\documentclass[pre,twocolumn]{revtex4}

\usepackage{amsmath}
\usepackage{amsfonts}
\usepackage{hyperref}
\usepackage{graphicx}

\usepackage[T1]{fontenc}

\newcommand{\dd}{\mathrm{d}}

\newcommand{\tauversore}{\hat{\boldsymbol{\tau}}}

\newcommand{\vettoreg}{\boldsymbol{g}}

\newcommand{\vettoreN}{\boldsymbol{N}}

\newcommand{\acceleration}{\boldsymbol{a}}
\newcommand{\velocity}{\boldsymbol{v}}

\begin{document}

\title{Huygens' cycloidal pendulum: an elementary derivation}

\author{Riccardo Borghi}

\affiliation{Dipartimento di Ingegneria, Universit\`a degli Studi ``Roma tre''\\
Via Vito Volterra 62, I-00146 Rome, Italy\\
Riccardo.Borghi@uniroma3.it}

\begin{abstract}
A pedagogical  derivation of the Huygens cycloidal pendulum, suitable for undergraduates, is here presented. 
Our derivation rests only on simple algebraic and geometrical tricks, without the need of Calculus. 
\end{abstract}

\maketitle

\section{Introduction}
\label{Sec:Introduction}

It is a well known fact that the motion of a simple pendulum is far from being rigorously
isochrone, being the period of its oscillations an increasing function of their amplitude~\cite{Berkeley/1973}. 
For ``small'' oscillations the swinging time is assumed to be  approximately independent of the amplitude,
and it is within such a regime that simple pendulums are employed as clocks.
The ``isochronism question'' was known since Galileo's time, and in 1659 Christiaan Huygens,
``the most ingenious watchmaker of all time'', to use Sommerfeld's words~\cite{Sommerfeld/1970},
was the first to find the solution~\cite{Bell/1947,Yoder/1988}. Instead of using a circular motion, the pendulum bob was forced to move across a \emph{cycloid}, i.e.,  
the trajectory drawn by a typical  point 
on the periphery of a bicycle wheel when the latter moves on the ground without slipping. 

While the simple pendulum is a central topic in any undergraduate and/or high-school physics course,
the same cannot be said as far as the cycloidal pendulum is concerned. The latter is rarely offered to 
students as an example of application of Newton's laws of motion, although it was studied already in Newton's \emph{Principia}~\cite{Gauld/2004}.
Most treatments of cycloidal pendulum rest on some concepts of differential geometry and Calculus which could not be yet available inside the 
math toolbox of first-year undergraduates. 
Cycloid is also  associated to the solution of the celebrated brachistochrone problem~\cite{Weisstein}, for which the principles of Calculus of Variation were first  developed~\cite{Sommerfeld/1970}. 
During last years a considerable deal of work has been done about elementary treatments of these ``magic'' properties 
of cycloid, namely brachistochronism and tautochronism (see for instance~\cite{Figueroa/Gutierrez/Fehr/1997,Brooks/Push/2002,Sawicki/2005,Boute/2012,Onorato/Mascheretti/DeAmbrosis/2013,Mungan/Lipscombe/2017,Ben-Abu/Wolfson/Eshach/Yizhaq/2018}).

In the present paper a pedagogical derivation of Huygens' pendulum isochronism is proposed. The isochronism property is derived from basic mechanical principles and 
by using only a trivial algebraic trick that, once 
translated into the vector language, leads to the very geometrical definition of cycloids without the need of Calculus. 

To help Teachers, what follows will be arranged as a whole didactical unit that could be proposed to undergraduates within a standard two-hour lecture.
Some mathematical steps could seem a little bit redundant, but this has been done  to keep the level as elementary as possible.

\section{The statement of the problem}
\label{Sec:SOP}

In Fig.~\ref{Fig:Geometry.1} the geometry of the problem is shown: 
\begin{figure}[!ht]
\centerline{\includegraphics[width=6cm,angle=0]{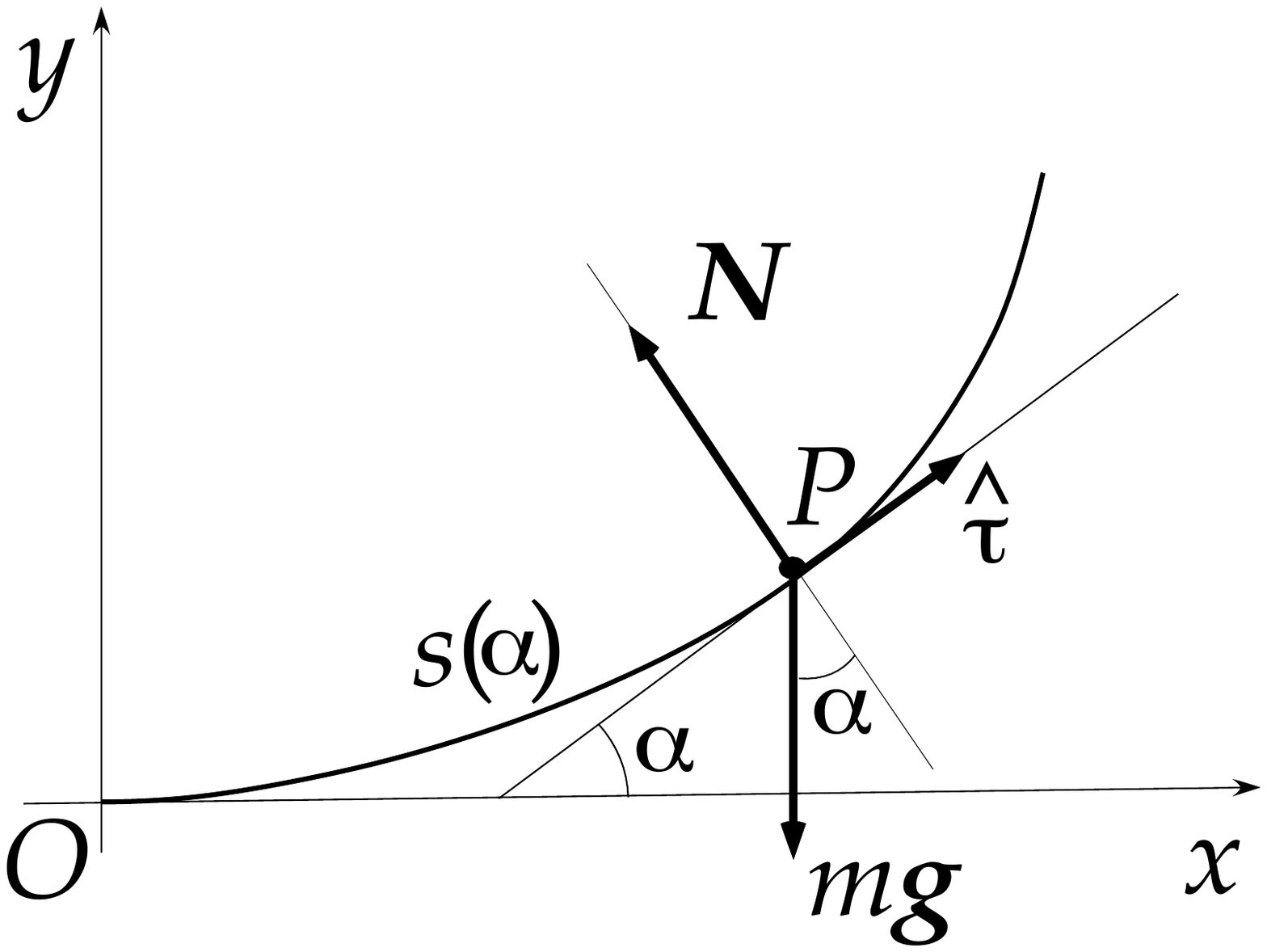}}
\caption{The isochrone pendulum geometry.} 
\label{Fig:Geometry.1}
\end{figure}

\noindent
$P$ is a point having mass $m$, sliding in a vertical plane $Oxy$  along a frictionless surface under the action of  its weight $m\vettoreg$. The vector $\vettoreN$
 describes the surface reaction which is directed along the surface normal, due to absence of friction.  
 Symbol $\tauversore$  denotes the  tangential unit vector. 
The task is to shape the surface in such a way the point motion be purely harmonic.   
To this end,  Newton's law is first written
\begin{equation}
\label{Eq:Equazione.1}
m\acceleration\,=\,m\vettoreg\,+\,\vettoreN\,,
\end{equation}
with $\acceleration$ being the point acceleration. Projection of both sides of Eq.~(\ref{Eq:Equazione.1})
along the tangential direction $\tauversore$  gives at once
\begin{equation}
\label{Eq:Equazione.2}
a_\tau\,=\,-g\,\sin\alpha\,,
\end{equation}
where symbols $a_\tau$ and $\alpha$ denote the tangential component of the acceleration and  the inclination angle of 
che curve with respect the horizontal direction $x$, respectively (see Fig.~\ref{Fig:Geometry.1}). 
Now, for  the motion along the trajectory to be rigorously harmonic it is mandatory that $a_\tau$
be proportional, up to a minus sign, to the curvilinear abscissa $s=\stackrel \frown {OP}$, 
\begin{equation}
\label{Eq:Equazione.3}
a_\tau\,=\,-\,\omega^2\,s\,,
\end{equation}
where $\omega=2\pi/T$, with $T$ denoting  the oscillation period.
On comparing Eqs.~(\ref{Eq:Equazione.2}) and~(\ref{Eq:Equazione.3}), the following equation is obtained:
\begin{equation}
\label{Eq:ImplicitDefinition}
s\,=\,\ell\,\sin\alpha\,,
\end{equation}
with  $\ell=g/\omega^2$.
Equation~(\ref{Eq:ImplicitDefinition})  represents an implicit definition of  
the trajectory shape in terms of the curvilinear 
abscissa $s$, the slope $\alpha$, and the characteristic length $\ell$.  

The task consists in proving that behind Eq.~(\ref{Eq:ImplicitDefinition})  a cycloid is hidden. 
To this end, it is worth first describing a ``standard solution'', based on a few concept of differential geometry.
This has been carried out in Sec.~\ref{Sec:StandardSolution}. 
Subsequently, our elementary solution will be presented in Sec.~\ref{Sec:ElementarySolution}.

\section{A ``standard'' solution}
\label{Sec:StandardSolution}

Our aim is to extract the parametric representation of the trajectory with respect to $\alpha$, i.e., functions $x=x(\alpha)$ and $y=y(\alpha)$.
To this end, both sides of Eq.~(\ref{Eq:ImplicitDefinition}) are first derived with respect to $\alpha$, 
\begin{equation}
\label{Eq:PendoliIsocronismo.4.1}
\begin{array}{l}
\displaystyle
\frac{\dd s}{\dd \alpha}\,=\,\ell\,\cos\alpha\,,
\end{array}
\end{equation}
where
\begin{equation}
\label{Eq:PendoliIsocronismo.4.1.2}
\begin{array}{l}
\displaystyle
\dd s\,=\,\sqrt{\dd x^2\,+\,\dd y^2}\,=\,
\dd x\,\sqrt{1+\tan^2\alpha}\,=\,\frac{\dd x}{\cos\alpha}\,.
\end{array}
\end{equation}
From Eqs.~(\ref{Eq:PendoliIsocronismo.4.1}) and~(\ref{Eq:PendoliIsocronismo.4.1.2}) we then have
\begin{equation}
\label{Eq:PendoliIsocronismo.4.2}
\begin{array}{l}
\displaystyle
\frac{\dd x}{\dd \alpha}\,=\,\frac{\dd x}{\dd s}\,\frac{\dd s}{\dd \alpha}\,=\,
\cos\alpha\,\frac{\dd s}{\dd \alpha}\,=\,\ell\,\cos^2\alpha\,,
\end{array}
\end{equation}
and, similarly for $y$,  
\begin{equation}
\label{Eq:PendoliIsocronismo.4.3}
\begin{array}{l}
\displaystyle
\frac{\dd y}{\dd \alpha}\,=\,\frac{\dd y}{\dd x}\,\frac{\dd x}{\dd \alpha}\,=\,
\tan\alpha\,\frac{\dd x}{\dd \alpha}\,=\,
\ell\,\sin\alpha\,\cos\alpha\,,
\end{array}
\end{equation}
where in the last step use has been made of Eq.~(\ref{Eq:PendoliIsocronismo.4.2}). On using elementary trigonometrics, the following system is then obtained: 
\begin{equation}
\label{Eq:PendoliIsocronismo.9}
\left\{
\begin{array}{l}
\displaystyle
\frac{\dd x}{\dd \alpha}\,=\,\frac{\ell}{2}\,(1\,+\,\cos\,2\alpha)\,,\\
\\
\displaystyle
\frac{\dd y}{\dd \alpha}\,=\,\frac{\ell}{2}\,\sin\,2\alpha\,.
\end{array}
\right.
\end{equation}
Finally, on introducing the parameter $R=\ell/4$ and on integrating the system~(\ref{Eq:PendoliIsocronismo.9}) with the initial condition $\{x(0),y(0)\}\equiv \{0,0\}$,
the required parametric representation of the trajectory is found
\begin{equation}
\label{Eq:PendoliIsocronismo.10}
\left\{
\begin{array}{l}
\displaystyle
x(\alpha)\,=\,R\,(2\alpha\,+\,\sin\,2\alpha)\,,\\
\\
\displaystyle
y(\alpha)\,=\,R\,(1\,-\,\cos\,2\alpha)\,,
\end{array}
\right.
\end{equation}
a cycloid, in fact.

\section{The elementary solution}
\label{Sec:ElementarySolution}

\begin{figure}[!ht]
\centerline{\includegraphics[width=6cm,angle=0]{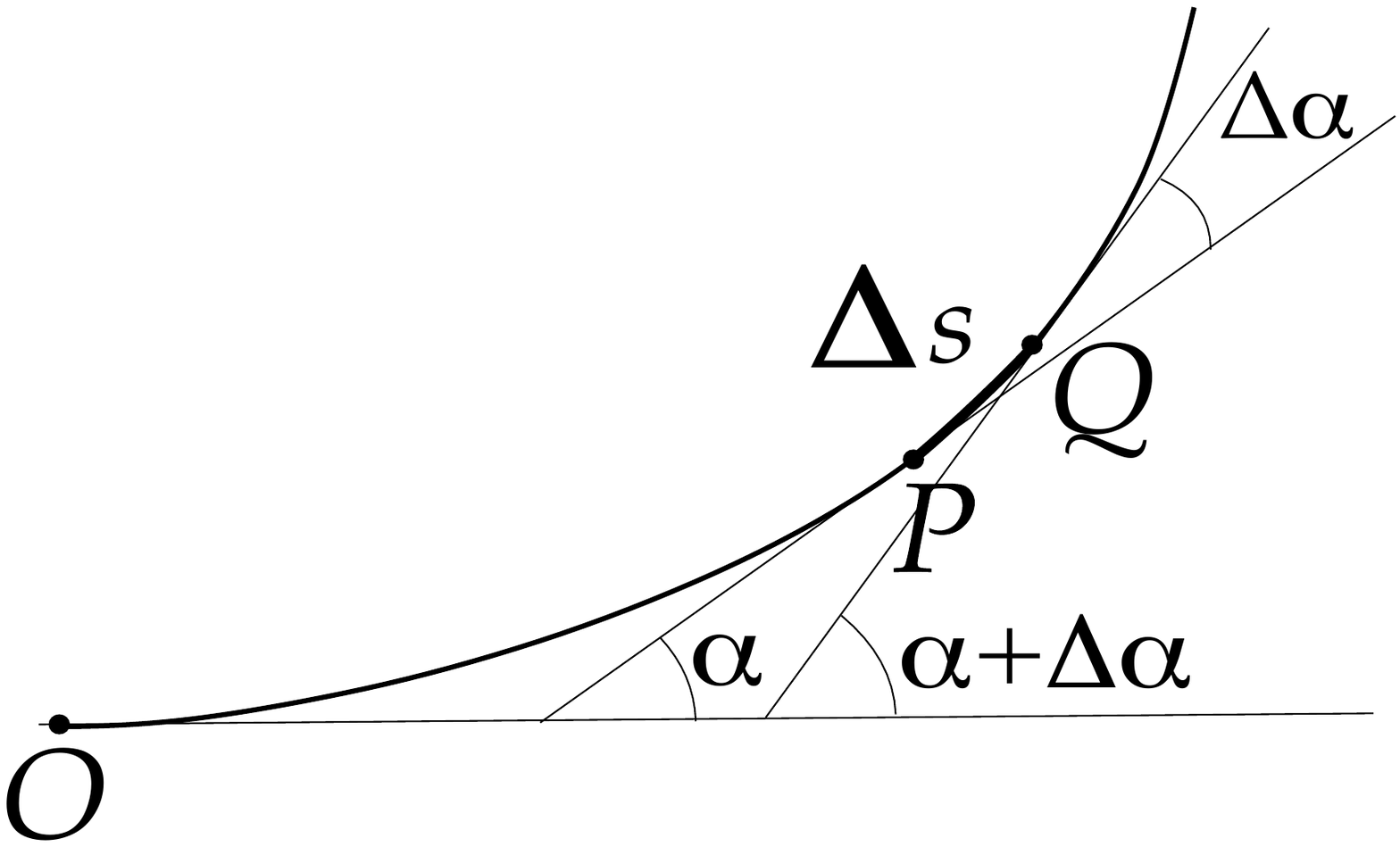}}
\caption{Geometrical interpretation of Eq.~(\ref{Eq:Equazione.3.1.1}).} 
\label{Fig:Geometry.1.1}
\end{figure}

The differential quantities $\mathrm{d}s$ and $\mathrm{d}\alpha$  are first replaced by finite, small quantities 
$\Delta s$ and $\Delta \alpha$, respectively. Equation~(\ref{Eq:PendoliIsocronismo.4.1}) can then be  introduced as follows:
consider a ``small'' displacement of the point $P$ to another  point along the curve, say $Q$, corresponding to the inclination
$\alpha+\Delta\alpha$, as sketched in Fig.~\ref{Fig:Geometry.1.1}. The displacement length, say $\Delta s$, coincides with the measure of the arc 
$\stackrel \frown {PQ}$, i.e.,
\begin{equation}
\label{Eq:Equazione.3.1.0.1}
\begin{array}{l}
\Delta s\,=\,{\stackrel \frown {OQ}}\,-\,{\stackrel \frown {OP}}\,=\,
\ell\,\left[\sin(\alpha\,+\,\Delta\alpha)\,-\,\sin\alpha\right]\,=\,\\
\\
\,=\,
\ell\,\left[\sin\alpha\,(1-\cos\Delta\alpha)\,-\,\cos\alpha\,\sin\Delta\alpha\right]\,,
\end{array}
\end{equation}
where use has been made of Eq.~(\ref{Eq:ImplicitDefinition}). 

For $Q$  sufficiently close to $P$,  i.e., for  $\Delta\alpha$  sufficiently small, the following approximations can be used:
\begin{equation}
\label{Eq:Equazione.3.1.0.1.0.1}
\left\{
\begin{array}{l}
\sin\Delta\alpha\simeq\Delta\alpha\,,\\
\\
\cos\Delta\alpha\simeq \,1\,,
\end{array}
\right.
\end{equation}
which, once inserted into Eq.~(\ref{Eq:Equazione.3.1.0.1}) leads to
\begin{equation}
\label{Eq:Equazione.3.1.1}
\Delta s\,=\,\ell\,\cos\alpha\,\Delta\alpha\,,
\end{equation}
and where the arc $\stackrel\frown {PQ}$  will be identified with the {segment} $\overline{PQ}$, as shown in Fig.~\ref{Fig:Geometry.1.1}.
Needless to say, Eq.~(\ref{Eq:Equazione.3.1.1}) is nothing but Eq.~(\ref{Eq:PendoliIsocronismo.4.1}).

\begin{figure}[!ht]
\centerline{\includegraphics[width=6cm,angle=0]{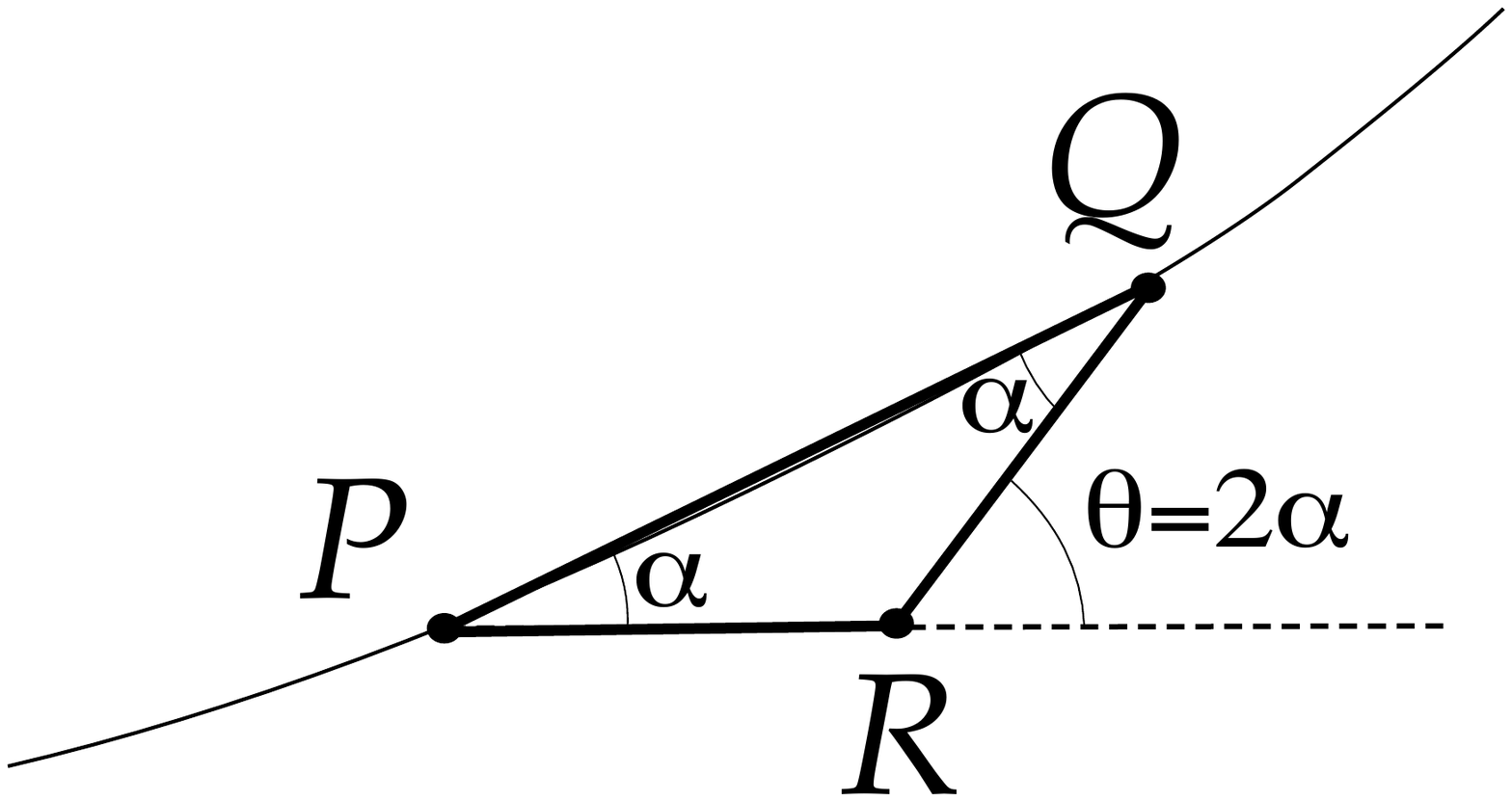}}
\caption{Geometrical interpretation of Eq.~(\ref{Eq:Equazione.3.1.1.1}).} 
\label{Fig:Geometry.2}
\end{figure}

The first algebraic trick consists to recast Eq.~(\ref{Eq:Equazione.3.1.1}) as follows:
\begin{equation}
\label{Eq:Equazione.3.1.1.1}
\Delta s\,=\,2\,\left(\dfrac \ell 2\,\Delta\alpha\right)\,\cos\alpha\,,
\end{equation}
and 
to introduce
the isoscele triangle  $PRQ$, built up in such a way that  $\overline{PR}\,=\,\overline{RQ}\,=\,\dfrac \ell 2\,\Delta\alpha$ (see Fig.~\ref{Fig:Geometry.2}).

On using simple geometrical considerations it follows that the horizontal inclination of $RQ$  is $\theta=2\alpha$. 
Accordingly, we have 
\begin{equation}
\label{Eq:Equazione.3.1.2.1}
\overline{PR}\,=\,\overline{RQ}\,=\,\dfrac \ell 4\,\Delta\theta\,,
\end{equation}
where it has been set $\Delta\theta=2\Delta\alpha$.
\begin{figure}[!ht]
\centerline{\includegraphics[width=6cm,angle=0]{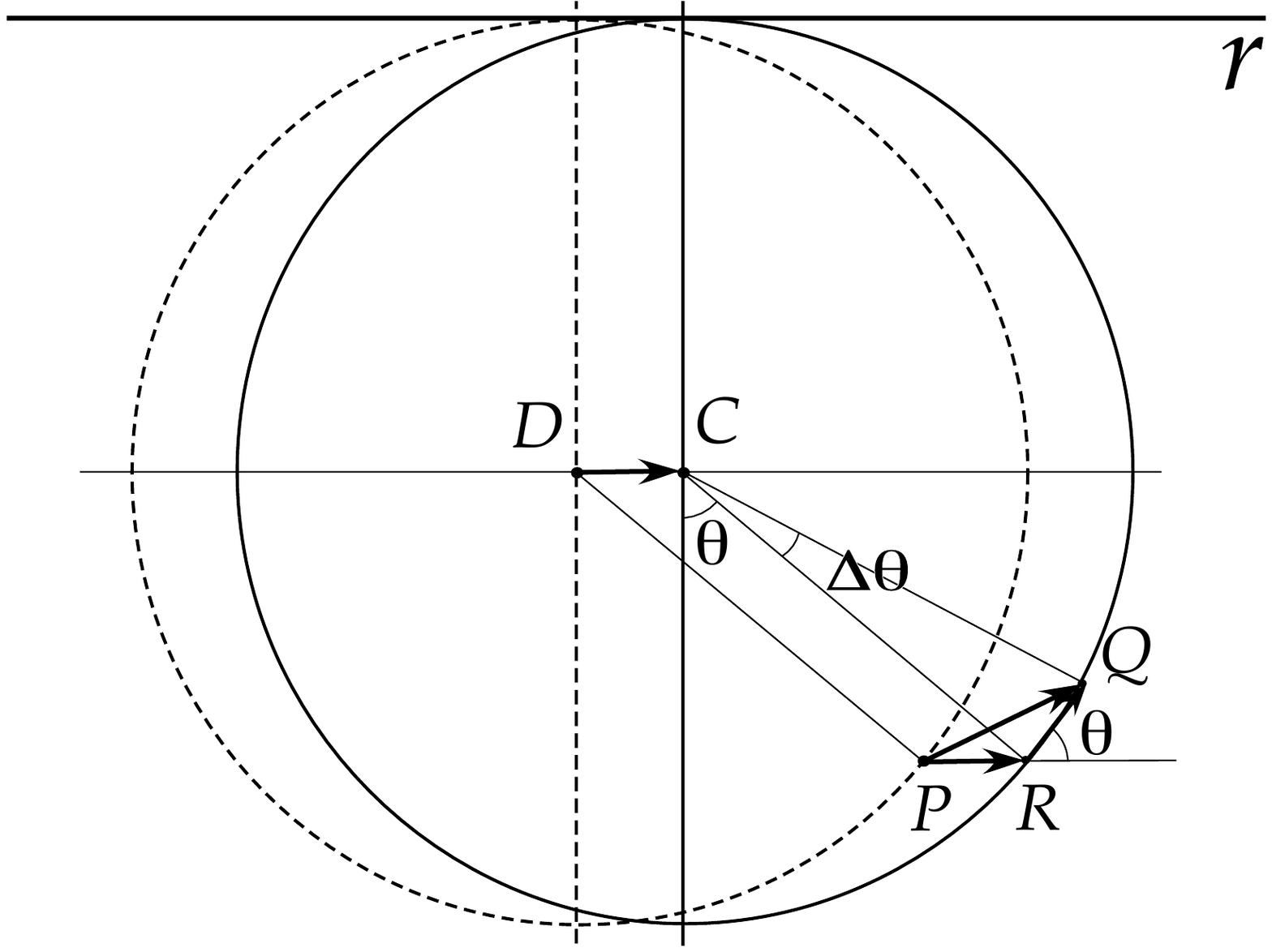}}
\caption{Geometrical interpretation of Eq.~(\ref{Eq:Equazione.3.1.2.1.1}) via a rolling circle.} 
\label{Fig:Geometry.3}
\end{figure}

Now the second trick: to give suitable orientations to the three sides of the triangle $PRQ$ in order for them to be interpreted as \emph{vectors} of the Euclidean plane.
This, in particular, allows the triangle itself to be represented through an algebraic equation, like for instance
\begin{equation}
\label{Eq:Equazione.3.1.2.1.1}
\overrightarrow{PQ}\,=\,\overrightarrow{PR}\,+\,\overrightarrow{RQ}\,,
\end{equation}
as sketched in Fig.~\ref{Fig:Geometry.3}. 
From  the same figure it also follows that the ratio 
${\overline{RQ}}/{\Delta\theta}\,=\,\ell /4$, can be interpreted as the radius of a circumference centred at the point $C$. 

Then, the final step: the vector $\overrightarrow{DC}=\overrightarrow{PR}$ is introduced in such a way  that 
Eq.~(\ref{Eq:Equazione.3.1.2.1.1})  is recast as follows:
\begin{equation}
\label{Eq:Equazione.3.1.2.1.1.1}
\overrightarrow{PQ}\,=\,\overrightarrow{DC}\,+\,\overrightarrow{RQ}\,.
\end{equation}

Accordingly, the total displacement $\overrightarrow{PQ}$  along the curve can be viewed as the result of the composition 
of two \emph{rigid} motions of the circumference: (i) an horizontal displacement represented by $\overrightarrow{DC}$,
plus  (ii)  a counterclockwise rotation by an angle  $\Delta\theta$ around the centre $C$. 
Furthermore, since $\overline{DC}=\overline{RQ}$, it should be immediately recognized that 
such a composition is equivalent to a perfect (i.e., without slipping) rolling of the circumference  along the horizontal line $r$, 
thus proving that the displacement $\overrightarrow{PQ}$ must  lay on a cycloid, Q.E.D.

\section{Conclusions: why is Huygens the ``most ingenious watchmaker of all time''?}
\label{Sec:CycloidEvolute}

\begin{figure}[!ht]
\centerline{\includegraphics[width=6cm,angle=-0]{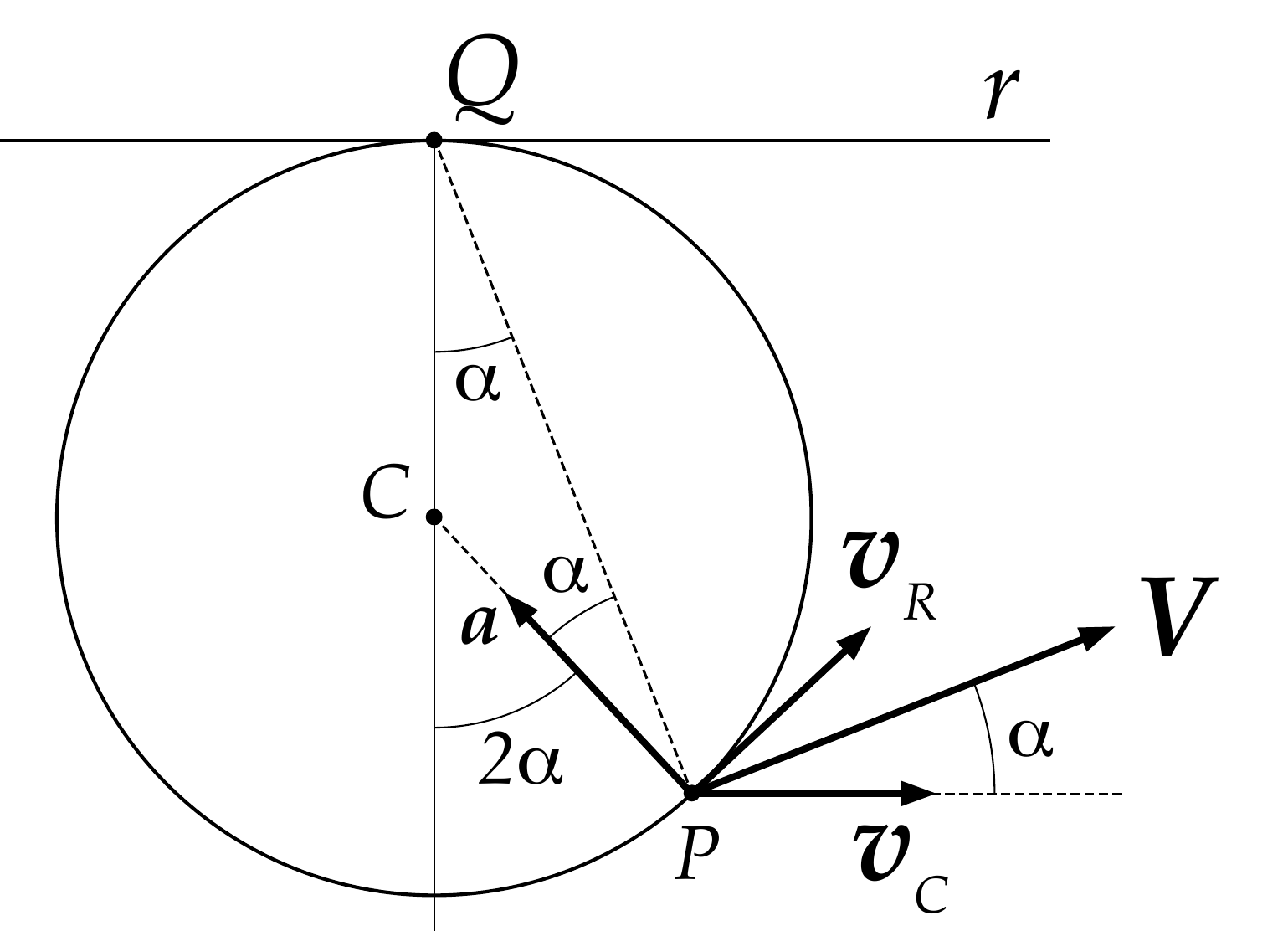}}
\caption{Cycloid's kinematics.} 
\label{Fig:CycloidEvolute.1}
\end{figure}

The Teacher could not resist to explain to his/her audience the Sommerfeld homage to Huygens. Again, this can be done  elementarly.
To this aim, the concept of \emph{evolute} of a curve, i.e., the geometric locus of its curvature centres, must first be introduced to students.
A possibility is to employ the kinematic strategy used, for instance, in~\cite{Borghi/2021} to evaluate the radius of curvature of conics. 
Accordingly, a point \emph{moving} across the cycloid is figured out, so that the radius of curvature follows from the classical Huygens formula involving normal 
component of acceleration and speed. 
For simplicity, the motion of the circle along the line $r$ will be supposed to be \emph{uniform}. In this way it is not difficult to realize that the velocity 
of $P$, say $\boldsymbol{V}$, can be thought of as the sum of two velocities: (i) that pertinent to the (uniform) rectilinear motion of the centre $C$, say $\velocity_C$, 
and (ii) the velocity of $P$, say $\velocity_R$, due to its uniform circular motion around $C$, as sketched into Fig.~\ref{Fig:CycloidEvolute.1}. 

Due to the absence of slipping along $r$ (pure rolling), it can be set $v_C=v_R=v$, so that the speed of $P$ turns out to be
\begin{equation}
\label{Eq:CycloidEvolute.1}
V\,=\,2\,v\,\cos\alpha\,.
\end{equation}
Moreover, since the motion of $P$ results from  the composition of a uniform rectilinear and a uniform circular ones, the point acceleration $\acceleration$ will be constantly directed toward the 
centre $C$, its modulus being 
\begin{equation}
\label{Eq:CycloidEvolute.1.2}
a\,=\,\dfrac{v^2}{\overline{PC}}\,.
\end{equation}
The normal component of the acceleration, say $a_\nu$, is then obtained by projecting $\acceleration$
along the direction $PQ$ of Fig.~\ref{Fig:CycloidEvolute.1}, which is by construction normal to the direction of $\boldsymbol{V}$,
\begin{equation}
\label{Eq:CycloidEvolute.2}
a_\nu\,=\,\dfrac{v^2}{\overline{PC}}\,\cos\alpha\,.
\end{equation}
On using Eqs.~(\ref{Eq:CycloidEvolute.1}) and~(\ref{Eq:CycloidEvolute.2}),  the radius of curvature of the cycloid at $P$, say $\rho$, is then obtained,
\begin{equation}
\label{Eq:CycloidEvolute.3}
\rho\,=\,\dfrac{V^2}{a_\nu}\,=\,4\,{\overline{PC}}\,\cos\alpha\,=\,2\,\overline{QP}\,,
\end{equation}
where in the last step use has been made of the fact that the triangle $CPQ$ is isoscele (see Fig.~\ref{Fig:CycloidEvolute.1}).
\begin{figure}[!ht]
\centerline{\includegraphics[width=4.5cm,angle=-0]{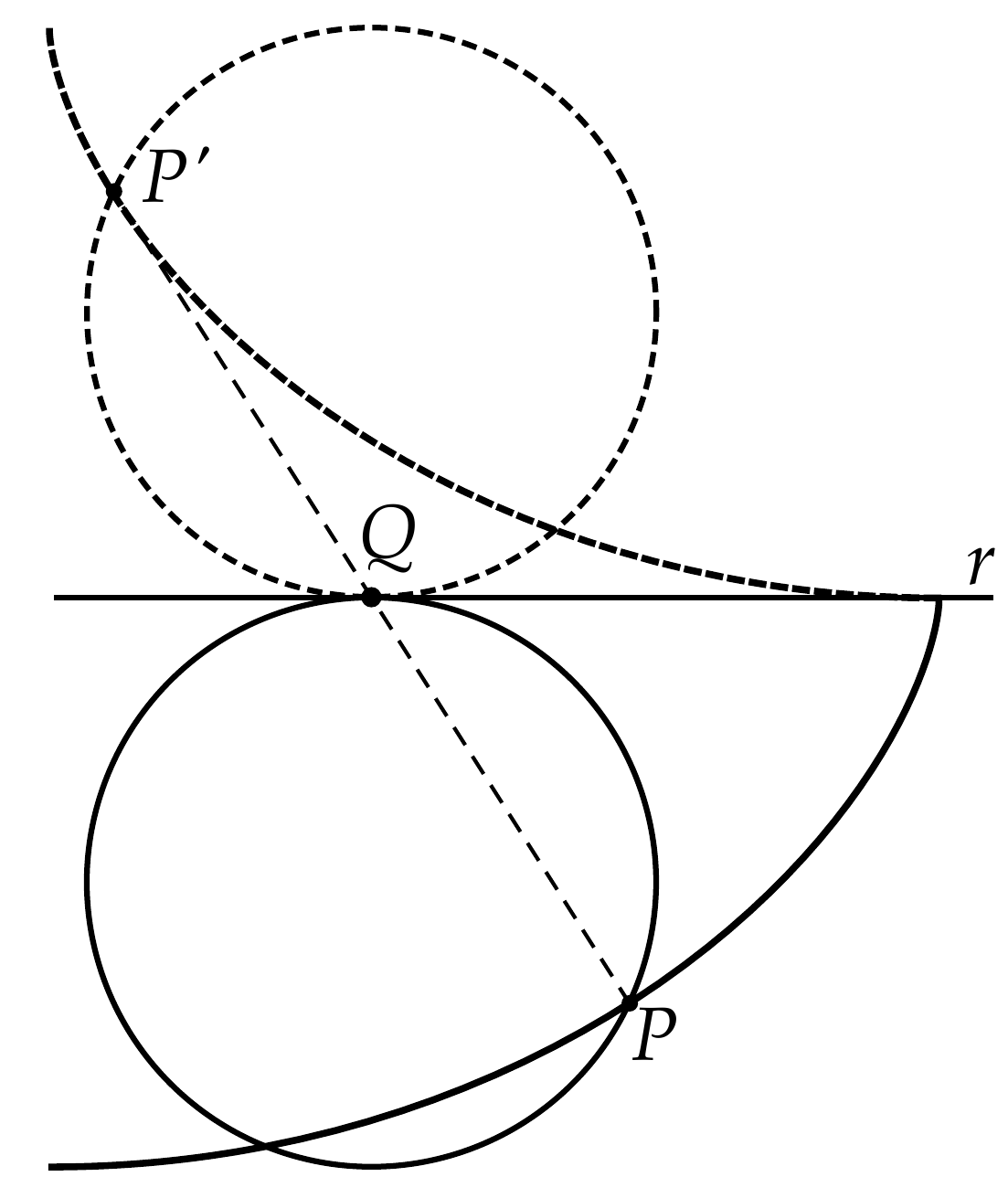}}
\caption{Cycloid's evolute is a cycloid.} 
\label{Fig:CycloidEvolute.2}
\end{figure}

\begin{figure}[!ht]
\centerline{\includegraphics[width=6cm,angle=-0]{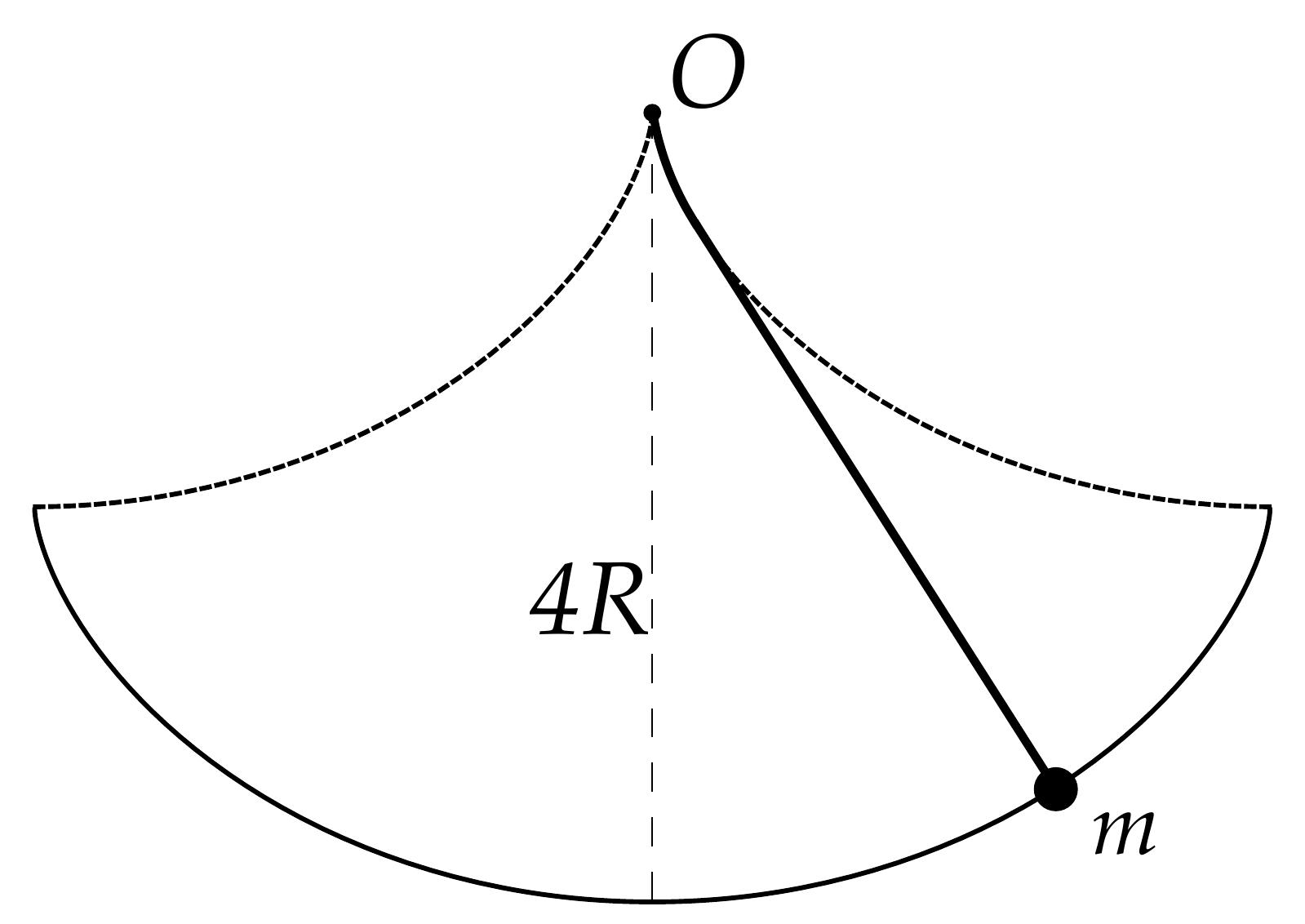}}
\caption{Huygens's isochronous cycolidal pendulum.} 
\label{Fig:CycloidEvolute.3}
\end{figure}

Equation~(\ref{Eq:CycloidEvolute.3})  implies that the centre of curvature, say $P'$, is placed symmetrically to $P$ with respect to the contact point $Q$ between the cycloid generatrix 
circle and $r$, as shown in Fig.~\ref{Fig:CycloidEvolute.2}. 
On introducing an identical circle (the dashed one in Fig.~\ref{Fig:CycloidEvolute.2}) in  the opposite side with respect $r$, it is not difficult to realize that also $P'$ must belong to an 
identical cycloid  (the dashed one in Fig.~\ref{Fig:CycloidEvolute.2}). 
Now, it could be proved that the direction $P'P$  is tangent to such cycloid (this could be left to our students as a useful exercise). 
This, together with all has been said and proved above, should be enough to help our students to grasp the way Huygens practically made his isochronous pendulum.
To this end, it is sufficient to suspend a point mass on an ideal, perfectly flexible and inextensible rope which is left bending along a restricting barrier having the dashed cycloidal shape,
as shown in Fig.~\ref{Fig:CycloidEvolute.3}.

To worthily conclude the present lecture, no words would be better than those employed in~\cite{Sommerfeld/1970} by Sommerfeld to masterfully 
describe  Huygens' genius:
\begin{quotation}
{\em Just as remarkable as Huygens' discovery of the isochronism of the cycloidal pendulum is the way in which he actually achieved the frictionless motion of the bob in the cycloid.
He availed himself of the rule that the evolute of a cycloid is another cycloid equal to the generating one. If, therefore, we tie a string of length $l=4R$ to the point $O$ of Fig.~\ref{Fig:CycloidEvolute.3}
in which the two upper cycloid arcs form a cusp, and if this string be pulled taut so that it rests against the right part of the cycloid (or the left part if deflected to the left), the endpoint $P$ of the string 
describes the lower cycloidal arc. The guiding of the bob along the lower cycloid effected in this manner is almost as frictionless as the guiding of the simple pendulum along a circular arc.}
\end{quotation}

\section*{Acknowledgments}

I wish to thank Turi Maria Spinozzi for his invaluable help during the preparation of the work.



\end{document}